\documentclass[aps,prc,reprint,superscriptaddress,nofootinbib]{revtex4-2}
\usepackage{amsmath}
\usepackage{amssymb}
\usepackage{graphicx}
\usepackage{dcolumn}
\usepackage{bm}
\usepackage{braket}
\usepackage{mathrsfs}
\usepackage{booktabs}
\usepackage{multirow}
\usepackage{url}
\usepackage{physics}
\usepackage{subfigure}
\usepackage[colorlinks=true,linkcolor=blue, filecolor=blue,urlcolor=blue,citecolor=blue]{hyperref}
\usepackage{color}

\begin{document}
\title
{ Ground-state properties of finite nuclei in relativistic Hartree-Bogoliubov theory with an improved quark mass density-dependent model}

\author{Renli Xu}
\email{renlixu@njucm.edu.cn}
\affiliation{School of Artificial Intelligence and Information Technology, Jiangsu Province Engineering Research Center of TCM Intelligence Health Service, Nanjing University of Chinese Medicine, Nanjing 210023, China}

\author{Chen Wu}
\email{wuchenoffd@gmail.com}
\affiliation{School of Physics and Electronic Information Engineering, Zhejiang Normal University, Jinhua 321004, China}

\author{Jian Liu}
\email{liujian@upc.edu.cn}
\affiliation{College of Science, China University of Petroleum (East China), Qingdao 266580, China}

\author{Bin Hong}
\affiliation{School of Physics Science and Engineering, Tongji University, Shanghai 200092, China}
\affiliation{Key Laboratory of Advanced Micro-Structure Materials, Ministry of Education, Shanghai 200092, China}

\author{Jie Peng}
\affiliation{School of Physics and Optoelectronic Engineering, Xiangtan University, Xiangtan 411105, China}

\author{Xiong Li}
\affiliation{School of Artificial Intelligence and Information Technology, Jiangsu Province Engineering Research Center of TCM Intelligence Health Service, Nanjing University of Chinese Medicine, Nanjing 210023, China}

\author{Ruxian Zhu}
\affiliation{School of Artificial Intelligence and Information Technology, Jiangsu Province Engineering Research Center of TCM Intelligence Health Service, Nanjing University of Chinese Medicine, Nanjing 210023, China}

\author{Zhizhen Zhao}
\affiliation{School of Artificial Intelligence and Information Technology, Jiangsu Province Engineering Research Center of TCM Intelligence Health Service, Nanjing University of Chinese Medicine, Nanjing 210023, China}

\author{Zhongzhou Ren}
\email{zren@tongji.edu.cn}
\affiliation{School of Physics Science and Engineering, Tongji University, Shanghai 200092, China}
\affiliation{Key Laboratory of Advanced Micro-Structure Materials, Ministry of Education, Shanghai 200092, China}

\begin{abstract}

A relativistic Hartree-Bogoliubov (RHB) model based on quark-meson coupling is developed, with a new parametrization derived from experimental observables. Using this model, we systematically investigate the ground-state properties of even-even nuclei spanning $8\leq Z\leq118$, including binding energies, quadrupole deformations, root-mean-square (rms) charge radii, two-nucleon separation energies, two-nucleon shell gaps, and $\alpha$-decay energies. Comparisons with available experimental data demonstrate that this subnucleon-based RHB model reliably describes the ground-state properties of finite nuclei.

\end{abstract}

\maketitle

\section{Introduction}

Nuclear energy density functionals (EDFs), grounded in density functional theory and the mean-field approach, offer a universally applicable and powerful framework for describing the properties of finite nuclei and infinite nuclear matter. Over the past several decades, a variety of EDFs have been proposed, which can be classified into two primary categories: non-relativistic and relativistic approaches. Within the non-relativistic domain, the most successful EDFs are the Hartree-Fock models based on density-dependent forces, such as the Skyrme force for zero-range interactions \cite{Vautherin1972,Vautherin1973} and the Gogny force for finite-range interactions \cite{Gogny1980}. As a relativistic version of the EDFs, the relativistic mean field (RMF) model describes nucleon interactions through the exchange of various virtual mesons \cite{Walecka1974,Serot1986,Horowitz1981,Gambhir1990,Sugahara1994,Ring1996,Lalazissis1997,Lalazissis1999,Bender2003,Piekarewicz2005,Lalazissis2009,Fattoyev2010}. Naturally incorporating relativistic effects, the RMF model treats spin-orbit coupling in a fully self-consistent manner, without requiring additional adjustable parameters. By fitting a few free parameters to experimental data, both non-relativistic and relativistic mean-field models enable an insightful understanding of a wide range of nuclear properties\cite{Schunck2008,Mizutori2000,Furnstahl2002,Reinhard2009,Sagawa1992,Ren1995,Ren1998,Ren1996,Grasso2006,Pei2006,Zhang2023,Zhou2010,Dobaczewski1996,Demetriou2001,Kruppa2000,Zheng2024,Reinhard1999,Vretenar1999,Sarazin2000,Ren2002,Quan2017,Ljungvall2008,Brown1998,Vretenar2011,Nazarewicz1996,Goriely2001,Stoitsov2003,Hilaire2007,Wang2023,WangHK2023,Zhang2025,Sheng2010}, including phenomena such as the nuclear skin and halo \cite{Schunck2008,Mizutori2000,Furnstahl2002,Reinhard2009,Sagawa1992,Ren1995,Ren1998,Ren1996,Grasso2006,Pei2006,Zhang2023,Zhou2010}, shell effects \cite{Dobaczewski1996,Demetriou2001,Kruppa2000,Zheng2024}, shape coexistence \cite{Reinhard1999,Vretenar1999,Sarazin2000,Ren2002,Quan2017,Ljungvall2008}, exotic nuclear structures \cite{Brown1998,Vretenar2011}, and the proton and neutron drip lines \cite{Nazarewicz1996,Goriely2001,Stoitsov2003,Hilaire2007}. These models also shed light on various properties of neutron stars \cite{Chabanat1997,Shen1998,Malik2018,Chen2014,Hong2024}.\par    

The RMF framework has evolved into a wide range of theoretical formulations to date. In the original Walecka model \cite{Walecka1974}, the nuclear interaction is described via RMF theory with nucleons interacting through the exchange of scalar ($\sigma$) and vector ($\omega$) mesons. Subsequently, Boguta and Bodmer extended the Walecka model by incorporating cubic and quartic self-interactions of the $\sigma$-meson, thereby enhancing the model's description of the nuclear matter incompressibility and surface characteristics \cite{Boguta1977}. The self-coupling and cross-coupling terms of the vector fields have also been systematically investigated \cite{Sugahara1994,Gmuca1992}. Notably, the introduction of isoscalar-isovector ($\omega$-$\rho$) coupling significantly influences the density dependence of the symmetry energy \cite{Piekarewicz2005,Horowitz2001,Horowitz2001-2,Reed2024}. Although the standard approach outlined above is widely used, several studies have extended it to explore modifications and alternatives. For instance, within the standard RMF framework, the derivative couplings of the scalar and vector fields have been proposed \cite{Zimanyi1990,Rufa1988}. Relative to the nonlinear meson self-interactions, relativistic models that explicitly incorporate density-dependent meson-nucleon couplings have seen significant advancements \cite{Brockmann1992,Lalazissis2005,Typel1999,Long2004,Vretenar2002}.
Besides these, the point-coupling model, which describes nucleon interactions via relativistic zero-range effective interactions, also marks a key progression within the RMF framework \cite{Nikolaus1992,Rusnak1997,Madland2002,Zhao2010}.\par

Alternatively, various effective models incorporating sub-nucleonic degrees of freedom have also been successfully developed. A prominent example is the quark-meson coupling (QMC) model, originally formulated by Guichon in 1988 \cite{Guichon1988}. This innovative approach couples the meson fields directly to the confined quarks within nucleons, establishing a theoretical framework that enables systematic investigation of nucleon modifications in the nuclear medium and provides critical insights into the fundamental properties of nuclear matter \cite{Fleck1990,Saito1994,Menezes2005,Guichon2018,Saito2007}. Other theoretical frameworks based on quark-meson degrees of freedom include the chiral SU(3) quark model \cite{Papazoglou1998,Papazoglou1999,Wang2003,Wang2001}, the quark mean field (QMF) model \cite{Toki1998,Shen2000,Hu2014}, and the improved quark mass density-dependent (IQMDD) model \cite{Wu2005,Wu2008,Wu2009}, among others. The IQMDD model is developed on the basis of the quark mass density-dependent (QMDD) model proposed by Fowler, Raha and Weiner \cite{Fowler1981}. According to the QMDD model, the masses of the up (u), down (d), and strange (s) quarks, as well as their corresponding antiquarks, are given by 
\begin{eqnarray}
	m_{q} & =&\frac{B}{3n_B}\quad(q=u,d,\bar{u},\bar{d}), \\
	m_{s,\bar{s}} & =&m_{s0}+\frac{B}{3n_B},
\end{eqnarray}
where $n_B$ denotes the baryon number density, $m_{s0}$ represents the current mass of the strange quark, and $B$ is the bag constant.
Since perturbative QCD fails to provide a confinement solution for quarks, a confinement potential, typically proportional to $r$ (or $r^2$), is introduced into the quark system to prevent quarks from reaching regions that are infinitely distant or excessively large \cite{Toki1998,Shen2000}. In the QMDD model, the confinement mechanism is manifested by requiring the mass of isolated quarks to approach infinity, rendering the vacuum incapable of supporting them \cite{Peng1999,Benvenuto1995}. As the volume approaches infinity or the density tends to zero, the quark mass diverges, which is precisely the behavior characterized by equations (1) and (2). This confinement mechanism is analogous to that in the MIT bag model \cite{Chodos1974,Chodos1974-2,Gilson1993}. Building upon the QMDD framework, the IQMDD model extends the original formalism by systematically incorporating $\sigma$, $\omega$, and $\rho$ meson fields. This generalization establishes a self-consistent coupling between quark degrees of freedom and mesonic fields throughout the nuclear medium, thereby enabling rigorous nuclear many-body calculations within the mean-field approximation \cite{Wu2010}.\par

On the other hand, the construction and operation of new-generation radioactive ion beam (RIB) facilities have produced an increasing number of nuclei far from $\beta$-stability valley, opening up new frontiers in nuclear physics that expand our understanding of nuclear phenomena from stable nuclei to exotic ones \cite{Zhan2010,Motobayashi2010,Thoennessen2010}. Exotic nuclei near the drip lines are weakly-bound systems, where the Fermi surface of protons (or neutrons) is close to the particle continuum. This leads to an increased scattering of Cooper pairs into the continuum states as a result of pairing correlations, and furthermore, the conventional BCS model with the monopole pairing force can only provide a poor approximation in this case \cite{Vretenar2005}. For weakly bound nuclei, a unified and self-consistent treatment of the mean field and pairing correlations becomes essential. The development of the RHB model has provided an effective approach to address this issue \cite{Vretenar1999,Meng1996,Vretenar1997,Tian2009,Geng2022}. In the RHB model, a unified description of particle-hole (ph) and particle-particle (pp) correlations is given, and the continuum effects are also effectively taken into account in the coordinate representation \cite{Zhang2020,Zhou2000,Zhang2013,Meng1998,Zhou2003,Zhang2022b}. In this paper, RHB calculations incorporating quark-meson coupling are presented within the IQMDD framework, serving as a natural extension of the IQMDD model. With this new model, we systematically calculated the ground-state properties of even-even nuclei with proton numbers ranging from $Z$=8 to $Z$=118 where experimental data are available. This paper is organized as follows. The main formulas of the IQMDD+RHB model are presented in Section 2. In Section 3, the model parameters and numerical results are presented and discussed. Finally, a summary is given in Section 4.\par

\section{Theoretical framework}\label{}
The effective Lagrangian density of the IQMDD model can be expressed as
\begin{eqnarray}
	\mathcal{L} &=&
	\bar{\varphi}_q\bigg\{\gamma^\mu\bigg[i\partial_\mu-g_\omega^q\omega_\mu-\frac{g_\rho^q}{2}\tau^q\cdot\rho_\mu-\frac{e}{2}\bigg(\frac{1}{3}+\tau_3^q\bigg)A_\mu\bigg]\nonumber\\&&
	+\frac{f_\omega g_\omega^q}{2M_N}\sigma^{\mu\nu}\partial_\nu\omega_\mu-m_q+g_\sigma^q\sigma\biggr\}\varphi_q+\frac{1}{2}\partial^\mu\sigma\partial_\mu\sigma-U(\sigma)\nonumber\\&&-\frac{1}{4}\Omega^{\mu\nu}\Omega_{\mu\nu}+\frac{1}{2}m_{\omega}^{2}\omega^{\mu}\omega_{\mu}-\frac{1}{4}G^{\mu\nu}G_{\mu\nu}+\frac{1}{2}m_{\rho}^{2}\rho^{\mu}\rho_{\mu}\nonumber\\&&-\frac{1}{4}F^{\mu\nu}F_{\mu\nu}.
\end{eqnarray}
The strength tensors for the vector mesons and the electromagnetic field are defined as follows: $\Omega_{\mu\nu}=\partial_\mu\omega_\nu-\partial_\nu\omega_\mu$, $G_{\mu\nu}=\partial_{\mu}\rho_{\nu}-\partial_{\nu}\rho_{\mu}$ and $F_{\mu\nu}=\partial_{\mu}A_{\nu}-\partial_{\nu}A_{\mu}$.
The quark mass $m_q$ ($q = u$, $d$), characterized by its density dependence, is given by equation (1). The self-interaction potential of the $\sigma$ field is expressed in the following form
\begin{eqnarray}
	U(\sigma) = \frac{1}{2} m_{\sigma}^2 \sigma^2 + \frac{1}{3} b \sigma^3 + \frac{1}{4} c \sigma^4 + B.
\end{eqnarray}
The bag constant $B$ is introduced such that $U(\sigma_{\nu})=0$, where $\sigma_{\nu}$ denotes the value of the sigma field at which the potential $U(\sigma)$ attains its absolute minimum. The effective quark mass $m_q^*$ is given by
\begin{eqnarray}
	m_q^* = m_q - g^q_\sigma \sigma.
\end{eqnarray}
In nuclear matter, three quarks form a soliton bag, and the effective nucleon mass is derived from the bag energy and is expressed as
\begin{eqnarray}
	M_N^* &=& \sum_q E_q \nonumber\\ &=& \sum_q \frac{4}{3} \pi R^3 \frac{\Gamma_q}{(2 \pi)^3} \int_0^{K_F^q} \sqrt{m_q^{*2} + k^2} \left( \frac{dN_q}{dk} \right) dk,\nonumber\\
\end{eqnarray}
where $K^q_F$ is the Fermi momentum of the quarks, $\Gamma_q$ denotes the quark degeneracy. $dN_q/dk$ is the density of states for various types of quarks in a confined space. The expressions for $dN_q/dk$ and $K^q_F$ used in this work can be found in Ref. \cite{Zhang2001}. The bag radius $R$ can be obtained from the equilibrium condition for the nucleon bag
\begin{eqnarray}
	\frac{\delta M_N^*}{\delta R} = 0.
\end{eqnarray}

Our focus now shifts to examining finite nuclei within the IQMDD model. Analogous to the QMC model \cite{Saito2007}, one can formulate a relativistic Lagrangian density at the hadronic level:

\begin{eqnarray}
	\mathcal{L}_{H} & =&\bar{\psi}\bigg[i\gamma^\mu\partial_\mu-M_N^*(\sigma)-g_\omega\gamma^\mu\omega_\mu-\frac{g_\rho}{2}\gamma^\mu\tau\cdot\rho_\mu \nonumber\\
	&&-\frac{e}{2}\gamma^{\mu}(1+\tau_{3})A_{\mu}+\frac{f_{\omega}g_{\omega}}{2M_{N}}\sigma^{\mu\nu}\partial_{\nu}\omega_{\mu}\bigg]\psi \nonumber\\
	&&+\frac{1}{2}\partial^{\mu}\sigma\partial_{\mu}\sigma-U(\sigma)-\frac{1}{4}\Omega^{\mu\nu}\Omega_{\mu\nu} \nonumber\\
	&&+\frac{1}{2}m_{\omega}^{2}\omega^{\mu}\omega_{\mu}-\frac{1}{4}G^{\mu\nu}G_{\mu\nu}+\frac{1}{2}m_{\rho}^{2}\rho^{\mu}\rho_{\mu} \nonumber\\
	&&-\frac{1}{4}F^{\mu\nu}F_{\mu\nu}.
\end{eqnarray}
Starting from the Lagrangian density (8), the RHB equation for the nucleons, obtained through a self-consistent treatment of the mean field and pairing correlations, is presented as \cite{Kucharek1991}
\begin{eqnarray}
	\begin{pmatrix}
		h_D-\lambda & \Delta \\
		-\Delta^* & -h_D^*+\lambda
	\end{pmatrix}
	\begin{pmatrix}
		U_k \\
		V_k
	\end{pmatrix}=E_k
	\begin{pmatrix}
		U_k \\
		V_k
	\end{pmatrix}. 
\end{eqnarray}
Here, $\lambda$ represents the chemical potential, and $E_k$ denotes the quasiparticle energy. The single-nucleon Dirac Hamiltonian, $h_D$, is given by
\begin{eqnarray}
	h_D(\boldsymbol{r})=\boldsymbol{\alpha}\cdot\boldsymbol{p}+V(\boldsymbol{r})+\beta M_N^*.
\end{eqnarray}
The vector potential reads
\begin{eqnarray}
	V(\boldsymbol{r})&=&g_{\omega}\omega_{0}(\boldsymbol{r})\boldsymbol{I}+\frac{g_{\rho}}{2}\tau_{3}\rho_{0}(\boldsymbol{r})\boldsymbol{I}-\frac{f_\omega g_\omega}{2M_N}i\gamma^i\partial_i\omega_0(\boldsymbol{r})\nonumber\\&&+\frac{e}{2}(1+\tau_3)A_0(\boldsymbol{r})\boldsymbol{I},
\end{eqnarray}
where $\boldsymbol{I}$ represents the identity matrix. Formally, the equations of motion for the mesons and photon are given by
\begin{eqnarray}
	(-\Delta + m_\sigma^2) \sigma(\boldsymbol{r}) &=& -\frac{\partial M_N^*}{\partial \sigma} \rho_s(\boldsymbol{r}) - b \sigma^2(\boldsymbol{r}) - c \sigma^3(\boldsymbol{r}), \nonumber\\
	(-\Delta + m_\omega^2) \omega_0(\boldsymbol{r}) &=& g_\omega \rho_\upsilon(\boldsymbol{r}) + \frac{f_\omega g_\omega}{2 M_N} \rho_0^T(\boldsymbol{r}), \nonumber\\
	(-\Delta + m_\rho^2) \rho_0(\boldsymbol{r}) &=& \frac{g_\rho}{2} \rho_3(\boldsymbol{r}), \nonumber\\
	-\Delta A_0(\boldsymbol{r}) &=& e \rho_p(\boldsymbol{r}).
\end{eqnarray}
Here, $\rho_s$, $\rho_\upsilon$, $\rho_0^T$, $\rho_3$ and $\rho_p$ represent the densities of scalar, vector, tensor, the third component of isovector, and proton, respectively. The relativistic pairing field $\Delta$ in Eq. (9) can be expressed as
\begin{eqnarray}
	\Delta_{n_1n_1'} &=& \frac{1}{2} \sum_{n_2n_2'} \langle n_1n_1'| V^{pp} | n_2 n_2' \rangle \kappa_{n_2n_2'}.
\end{eqnarray}
The pairing tensor is written as
\begin{eqnarray}
	\kappa_{nn'} = \sum_{E_k > 0} V_{nk}^* U_{n'k}.
\end{eqnarray}
In the two-body pairing channel, the finite range force is used, and its form in coordinate space is given by \cite{Paar2014},
\begin{eqnarray}
	V^{pp}(\boldsymbol{r}_1, \boldsymbol{r}_2, \boldsymbol{r}_1', \boldsymbol{r}_2') &=& -G \delta(\boldsymbol{R} - \boldsymbol{R}') P(\boldsymbol{r}) P(\boldsymbol{r}'),\nonumber\\
	P(\boldsymbol{r}) &=& \frac{1}{(4\pi a^2)^{3/2}} e^{-\boldsymbol{r}^2 / 2a^2},
\end{eqnarray}
where $\boldsymbol{r}$ and $\boldsymbol{R}$ stand for the relative and center-of-mass coordinates of the two-nucleon system, respectively. The values of parameters $G$ and $a$ are adopted from Ref. \cite{Paar2014}.

The coupled equations mentioned above can be solved through a self-consistent iterative scheme, where the wave functions are expanded in terms of the harmonic oscillator basis functions. In our calculation, the number of oscillator shells is chosen as $N_f=12$ and $N_b=20$. From this solution, we can calculate the total binding energy, which has the following form:
\begin{eqnarray}
B(A,Z) &=&\mathrm{Tr}[h_D \rho] +E_{\mathrm{pair}}+E_{\mathrm{c.m.}} -AM_N\nonumber\\
&&-\frac{1}{2} \int \mathrm{d}^3\boldsymbol{r}\frac{\partial M_N^*}{\partial \sigma} \rho_s(\boldsymbol{r})\sigma(\boldsymbol{r})-\frac{1}{6} \int \mathrm{d}^3\boldsymbol{r}b \sigma^3(\boldsymbol{r})\nonumber\\
&&-\frac{1}{4} \int \mathrm{d}^3\boldsymbol{r}c \sigma^4(\boldsymbol{r}) - \frac{1}{2} \int \mathrm{d}^3\boldsymbol{r} g_\omega \rho_v(\boldsymbol{r}) \omega_0(\boldsymbol{r}) \nonumber\\
&&-\frac{1}{2} \int \mathrm{d}^3\boldsymbol{r} \left[ \frac{g_\rho}{2} \rho_3(\boldsymbol{r}) \rho_0(\boldsymbol{r}) + e \rho_p(\boldsymbol{r}) A_0(\boldsymbol{r}) \right] \nonumber\\
&&-\frac{1}{2} \int \mathrm{d}^3\boldsymbol{r} \frac{f_\omega g_\omega}{2 M_N} \rho_0^T(\boldsymbol{r}) \omega_0(\boldsymbol{r}). 
\end{eqnarray}
Here, $E_{\mathrm{pair}}$ denotes the nuclear ground-state pairing energy, which is given by
\begin{eqnarray}
	E_{\mathrm{pair}} = \frac{1}{4} \sum_{n_1 n_1'} \sum_{n_2 n_2'} \kappa_{n_1 n_1'}^* \langle n_1 n_1' | V^{pp} | n_2 n_2' \rangle \kappa_{n_2 n_2'}.
\end{eqnarray}
The microscopic center-of-mass correction, which is employed in our calculations, is given as:
\begin{eqnarray}
	E_{\mathrm{c.m.}} = -\frac{P_{\mathrm{c.m.}}^2}{2AM_N},
\end{eqnarray}
where $P_{\mathrm{c.m.}}$ denotes the total momentum as viewed from the center-of-mass frame.
\section{Numerical results and discussions}

\begin{table*}[htbp] \centering
	\begin{ruledtabular}
		\caption{Model parameters of the effective interactions IQMDD3 and IQMDD-2*. The parameter $b$ and the masses of the mesons $m_\sigma$, $m_\omega$ and $m_\rho$ are measured in MeV.}
		\label{tab:1}
		\begin{tabular}{cccccccccc} 
			&$m_\sigma$&$m_\omega$&$m_\rho$&$g^q_\sigma$&${g^q_\omega}$&${g^q_\rho}$
			&$b$&$f_\omega$&$\eta$\\ 
			\midrule
			IQMDD-2*&508.0&782.5&763.0&5.09&2.9796&9.7001&--3400.0&2.1&0.03\\
			IQMDD3&498.5998&782.5&763.0&4.9667&2.9404&9.2314&--3114.1540&2.0993&0.0\\
		\end{tabular}
	\end{ruledtabular}
\end{table*}

Prior to conducting the numerical calculations, we first examine the parameters of this model. The parameter set IQMDD-2*, incorporating both the quark-$\omega$ tensor coupling and the non-linear $\omega$-$\rho$ interaction term, has been implemented to investigate the properties of finite nuclear systems \cite{Xu2013}. This parameter set was initially proposed a decade ago, and subsequent experimental advancements have yielded substantial new data on nuclear mass and neutron skin thickness measurements. To explore potential improvements in finite nuclear property predictions, we conduct a global fitting procedure that combines the simulated annealing algorithm with the Levenberg-Marquardt method, resulting in the parameterization referred to as IQMDD3. The fitted nuclear properties include the binding energies and charge radii of twenty even-even nuclei: $^{16}$O, $^{40}$Ca, $^{48}$Ca, $^{58}$Ni, $^{86}$Kr, $^{88}$Sr, $^{90}$Zr, $^{92}$Mo, $^{116}$Sn, $^{124}$Sn, $^{132}$Sn, $^{134}$Te, $^{138}$Ba, $^{142}$Nd, $^{144}$Sm, $^{148}$Gd, $^{186}$Pb, $^{208}$Pb, $^{210}$Pb and $^{210}$Po. Spherical RHB calculations are employed in the fitting process. The experimental binding energy values are adopted from the atomic mass evaluation \cite{Wang2021}, and charge radii from Ref. \cite{Angeli2013}.

\begin{table}
	\begin{ruledtabular}
		\caption{Total binding energies B.E., charge radii $R_c$, and neutron-skin thickness $R_{np}$($R_{np}$=$R_{n}-R_{p}$) calculated with IQMDD3, compared to experimental data \cite{Wang2021,Angeli2013,Tao2021} (values in parentheses).}\label{tab:2}
		\begin{tabular}{cccc}
			Nucleus&B.E. (MeV)&$R_c$ (fm)&$R_{np}$ (fm)\\
			
			\midrule
			$^{16}$O&128.328 (127.619)&2.722 (2.699)&-0.03\\
			$^{40}$Ca&341.070 (342.052)&3.464 (3.478)&-0.05\\
			$^{48}$Ca&416.619 (416.001)&3.493 (3.477)&0.20\\
			$^{50}$Ti &438.112 (437.786)&3.574 (3.570)&0.13\\
			$^{52}$Cr &455.211 (456.352)&3.656 (3.645)&0.06\\
			$^{54}$Fe &469.653 (471.765)&3.704 (3.693)&0.00\\
			$^{58}$Ni &504.782 (506.460)&3.782 (3.776) &0.00\\
			$^{68}$Ni &590.678 (590.408)&3.899 (3.892$^a$) &0.21\\
			$^{86}$Kr &751.827 (749.235)&4.189 (4.184) &0.17\\
			$^{88}$Sr &770.741 (768.468)&4.235 (4.224) &0.13\\
			$^{90}$Zr &786.180 (783.897)&4.274 (4.269) &0.10\\
			$^{92}$Mo &799.110 (796.511)&4.319 (4.315) &0.06\\ 
			$^{112}$Sn &953.695 (953.525)&4.583 (4.595) &0.09\\
			$^{116}$Sn &989.254 (988.682)&4.616 (4.625) &0.14\\
			$^{124}$Sn &1051.822 (1049.958)&4.671 (4.674) &0.22\\
			$^{132}$Sn &1104.904 (1102.843)&4.728 (4.732) &0.30\\
			$^{134}$Te &1126.579 (1123.408)&4.765 (4.757) &0.26\\
			$^{136}$Xe &1145.636 (1141.882)&4.805 (4.796) &0.23\\
			$^{138}$Ba &1162.421 (1158.292)&4.843 (4.838) &0.20\\
			$^{140}$Ce &1176.638 (1172.683)&4.880 (4.877) &0.17\\
			$^{142}$Nd &1188.969 (1185.136)&4.915 (4.912) &0.14\\
			$^{144}$Sm &1199.256 (1195.730)&4.953 (4.952) &0.12\\
			$^{148}$Gd &1222.693 (1220.754)&5.003 (5.008) &0.11\\
			$^{184}$Pb &1435.552 (1432.022)&5.386 (5.393) &0.06\\
			$^{186}$Pb &1454.809 (1451.793)&5.397 (5.403) &0.08\\
			$^{204}$Pb &1608.613 (1607.506)&5.491 (5.480) &0.20\\
			$^{208}$Pb &1636.544 (1636.430)&5.512 (5.501) &0.24\\
			$^{210}$Pb &1645.453 (1645.553)&5.528 (5.521) &0.25\\
			$^{214}$Pb &1661.461 (1663.293)&5.562 (5.558) &0.27\\
			$^{210}$Po &1647.514 (1645.213)&5.548 (5.570) &0.21\\
		\end{tabular} 
	\end{ruledtabular}
	\raggedright
	$^a$Available experimental data taken from \cite{Tao2021};the remaining experimental charge radii are from \cite{Angeli2013}. 
\end{table}
	
The interaction strengths between the nucleon and the $\omega$- and $\rho$-mesons are selected as: $g_{\omega}=3g_{\omega}^q$ and $g_{\rho}=g_{\rho}^q$ \cite{Saito1994}. The bag constant is set to $B=174.0 ~\mathrm{MeV}\cdot\mathrm{fm}^{-3}$. The optimized parameters obtained for the IQMDD3 Lagrangian parameterization are presented in Table 1, with comparative results from the previous IQMDD-2*. To achieve better agreement with experimental neutron skin thickness measurements of $^{208}$Pb, the $\omega$-$\rho$ coupling constant ($\eta$) is set to zero in the IQMDD3 parameterization. Recent parity-violating electron scattering experiments at JLab have achieved a precision measurement of the neutron skin thickness in $^{208}$Pb, yielding a result of $R_n-R_p=0.283\pm0.071$ fm \cite{Adhikari2021}. Using dispersive optical model (DOM) analysis, Ref. \cite{Pruitt2020} determined the neutron skin thickness of $^{208}$Pb to be $0.12-0.25$ fm. Additional experimental results for this quantity are presented in Fig. 12 of Ref. \cite{Zhao2022}. The calculated value for $^{208}$Pb from the IQMDD3 model is $R_n-R_p=0.24$ fm, which falls within the experimental uncertainty range reported in Refs.\cite{Adhikari2021,Pruitt2020}. In Table 2, we present the IQMDD3 predictions for the total binding energies, charge radii, and neutron-skin thickness of the nuclei included in the fit, along with the calculated results for other isotopic chains. All calculations are performed within a spherical RHB approach. For the twenty nuclei used in the fitting, the rms deviations from IQMDD3 are 0.023 MeV for the binding energy per nucleon, 2.33 MeV for the total binding energy, and 0.01 fm for the charge radius.
	
Table 3 presents the spin-orbit splittings for $^{16}$O, $^{40}$Ca, $^{48}$Ca, $^{56}$Ni, $^{78}$Ni, $^{100}$Sn, $^{132}$Sn and $^{208}$Pb calculated with the DD-ME2, IQMDD-2*, and IQMDD3 parameter sets. The inclusion of tensor coupling significantly impacts the IQMDD model's description of nuclear spin-orbit splittings. As the tensor coupling strengths ( $g^q_\omega$ and $f_\omega$) of the IQMDD3 parameter set are slightly lower than those of IQMDD-2*, the calculated spin-orbit splittings for IQMDD3 are overall smaller than those for IQMDD-2*, as indicated in Table 3.
	
\begin{table}
	\begin{ruledtabular}
		\caption{Spin-orbit splittings (in MeV) of proton ($\pi$) and neutron ($\nu$) levels in closed-shell Nuclei: Calculations with DD-ME2\cite{Lalazissis2005}, IQMDD-2*, and IQMDD3 models compared to experimental data from Ref. \cite{Grawe2007}}\label{tab:3}
		\begin{tabular}{cccccc}
			Nucleus&State&DD-ME2&IQMDD-2*&IQMDD3&Expt.\\
			
			\midrule
			$^{16}$O&$\pi$(1p)&6.48&5.81&5.52&6.32\\
			&$\nu$(1p)&6.55&5.86&5.56&6.18\\	
			$^{40}$Ca&$\pi$(1d)&6.70&6.26&5.97&5.40\\
			&$\nu$(1d)&6.77&6.29&6.00&5.63\\
			$^{48}$Ca&$\pi$(2p)&1.56&2.25&2.16&1.50\\
			&$\nu$(2p)&1.51&1.65&1.57&2.03\\
			$^{56}$Ni&$\pi$(2p)&1.24&1.60&1.53&1.11\\
			&$\nu$(2p)&1.39&1.78&1.70&1.11\\			
			$^{78}$Ni&$\pi$(2p)&1.31&1.72&1.63&1.40\\
			&$\nu$(2p)&1.36&1.60&1.50&1.33\\			
			$^{100}$Sn&$\pi$(2p)&1.41&1.61&1.54&2.85\\
			&$\nu$(2d)&1.95&2.31&2.21&1.65\\			
			$^{132}$Sn&$\pi$(2d)&1.84&2.10&1.99&1.74\\
			&$\nu$(2d)&1.92&2.08&1.96&1.65\\			
			$^{208}$Pb&$\pi$(2d)&1.68&1.73&1.65&1.34\\
			&$\nu$(2g)&2.34&2.47&2.34&2.49\\
		\end{tabular} 
	\end{ruledtabular} 
\end{table}

Additionally, we present the findings of the bulk characteristics for infinite nuclear matter in Table 4. It presents corrected predictions for nuclear matter properties based on the IQMDD-2* parameter set as reported in Ref. \cite{Xu2013}. Compared with IQMDD-2*, the IQMDD3 parameterization, which lacks isoscalar-isovector coupling, yields a larger symmetry energy at saturation density. For the six parametrizations in Table 4, the neutron-skin thickness of $^{208}$Pb increases monotonically from 0.160 fm to 0.285 fm as the symmetry energy slope increases from 47.2 MeV to 122.6 MeV. This positive correlation, previously highlighted by the literature \cite{Roca-Maza2011}, indicates that a larger symmetry energy slope $L$ implies higher neutron matter pressure and a thicker neutron-skin thickness in $^{208}$Pb.

\begin{table}
	\begin{ruledtabular}
		\caption{Bulk properties for infinite nuclear matter at saturation density $\rho_0$ (in fm$^{-3}$) across different models \cite{Lalazissis1997,Lalazissis2009,Piekarewicz2005,Fattoyev2010}. The quantities $\varepsilon_0$, $S$, $L$ and $K$ (all in MeV) represent the binding energy per nucleon, the symmetry energy, the slope of the symmetry energy and incompressibility coefficient of symmetric nuclear matter at saturation density. The neutron-skin thickness of $^{208}$Pb, represented by $R_{np}$ (in fm), is also included.}\label{tab:4}
		\begin{tabular}{ccccccc}
			Model&$\rho_0$&$\varepsilon_0$&$S$&$L$&$K$&$R_{np}$\\
			\midrule
			IU-FSU&0.155&-16.40&31.30&47.2&231.3&0.160\\
			FSU&0.148&-16.30&32.59&60.5&230.0&0.205\\
			IQMDD-2*&0.155&-16.38&34.16&72.8&283.3&0.212\\
			IQMDD3&0.153&-16.41&36.00&100.1&292.8&0.236\\
			NL3&0.148&-16.24&37.40&118.5&271.8&0.277\\
			NL3*&0.150&-16.31&38.68&122.6&258.1&0.285\\
		\end{tabular} 
	\end{ruledtabular} 
\end{table}

\subsection{\label{sec:leve31}Binding energies}
\begin{figure}[htbp]
	\centering
	\includegraphics[width=0.49\textwidth]{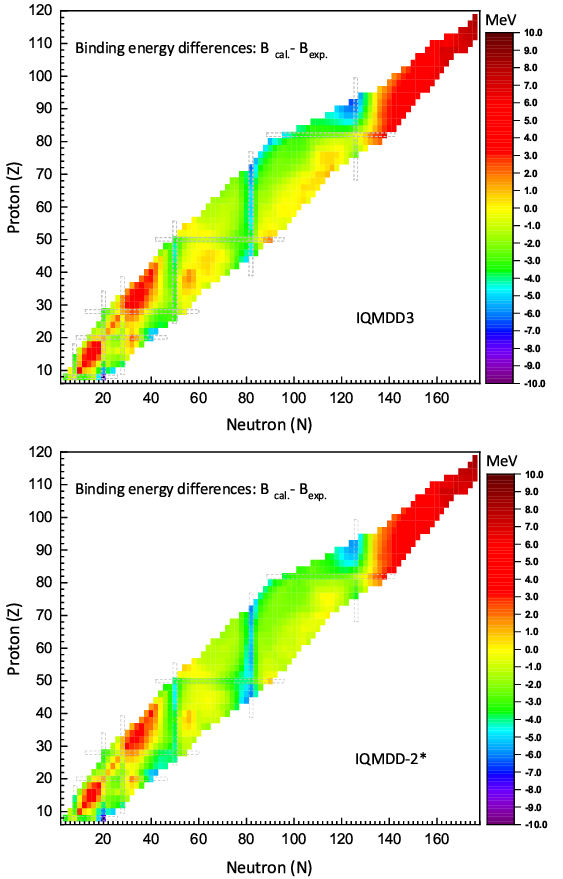}
	\caption{Differences in total binding energy between the calculations and experimental data for even-even nuclei using the IQMDD3 and IQMDD-2*. To guide the eye, the dashed frames correspond to the positions of the magic numbers 8, 20, 28, 50, 82, and 126.}
	\label{fig:1}
\end{figure}
We conduct systematic calculations for even-even isotopes from Oxygen ($Z$=8) to Oganesson ($Z$=118), focusing on nuclei with available experimental data. The differences in the total binding energy between calculated and experimental values for even-even nuclei spanning $Z$=8–118 are systematically presented in Fig.~\ref{fig:1}. The relevant experimental data are obtained from AME2020 atomic mass evaluation \cite{Wang2021}. For the specified mass region, our calculations encompass all 868 even-even nuclei in AME2020, containing 646 experimentally measured and 222 estimated mass values. Note that the binding energies herein and all subsequent results are obtained from axially deformed RHB calculations. As shown in Fig.~\ref{fig:1}, IQMDD3 and IQMDD-2* exhibit similar overall performance for total binding energy. In the fitting process, the binding energy per nucleon was employed as a constraint instead of the total binding energy. This adjustment effectively increased the fitting weight for light nuclei but led to a moderately larger deviation in the total binding energy of heavy nuclei compared to experimental values. This is reflected in the relatively large deviation in the total binding energy for heavy nuclei shown in Fig.~\ref{fig:1}. Statistical analysis reveals that the rms deviation of binding energy per nucleon is $\sigma_{B/A}=0.034$ MeV for IQMDD3 and $\sigma_{B/A}=0.036$ MeV for IQMDD-2*. For total binding energy, the corresponding values are $\sigma_{B}=2.893$ MeV for IQMDD3 and $\sigma_{B}=2.984$ MeV for IQMDD-2*. Additionally, statistical analysis shows that both models perform better in the medium and heavy nucleus regions compared to the light nucleus region in terms of the average binding energy per nucleon. When the analysis is confined to the range from Calcium ($Z$=20) to Oganesson ($Z$=118), the rms deviation decreases significantly to $\sigma_{B/A} =0.020$ MeV for IQMDD3 and $\sigma_{B/A} =0.022$ MeV for IQMDD-2*. 

For comparison, the typical binding energy rms deviations of established EDFs are 1.518 MeV for PC-PK1 by including the rotational correction energies \cite{Zhang2022}, 1.74 MeV for QMC$\pi$-$\mathrm{III}$-T \cite{Martinez2020}, 2.96 MeV for NL3* \cite{Agbemava2014}, 2.39 MeV for DD-ME2 \cite{Agbemava2014}, 2.29 MeV for DD-ME$\delta$ \cite{Agbemava2014}, 2.01 MeV for DD-PC1 \cite{Agbemava2014}, and 1.339 MeV for 91 spherical nuclei for PC-L3R \cite{Liu2023}. In terms of predicting nuclear binding energies, the performance of IQMDD3 is overall comparable to that of NL3*.

\subsection{\label{sec:leve32}Quadrupole deformations}
\begin{figure}[htbp]
	\centering
	\includegraphics[width=0.49\textwidth]{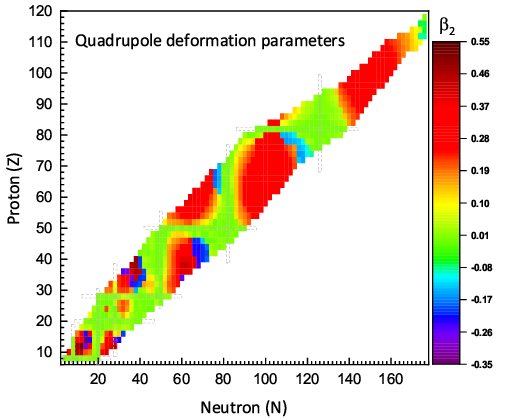}
	\caption{Quadrupole deformation parameters for even-even nuclei with proton numbers $8\leq Z\leq118$, calculated using the RHB approach with the IQMDD3 interaction.}
	\label{fig:2}
\end{figure}
Fig.~\ref{fig:2} illustrates the theoretical predictions of ground-state quadrupole deformation parameters $\beta_2$ for all the isotopic chains examined in this work. The deformation parameter $\beta_2$ is derived from the calculated nucleon quadrupole moments $Q$ via \cite{Gambhir1990}:
\begin{eqnarray}
	Q = \sqrt{\frac{16\pi}{5}} \frac{3}{4\pi} A R_0^2 \beta_2,
\end{eqnarray}
with $R_0=1.2A^{1/3}$(fm). The $\beta_2$ parameter, a dimensionless quantity with an absolute value less than 1, is characterized by relatively large empirical uncertainties. Given that our primary objective is to investigate the systematic trends and global characteristics of nuclear deformation patterns predicted by the IQMDD3 model, a comparison with experimental values is not included in the present analysis. As illustrated in Fig.~\ref{fig:2}, nuclei with spherical configurations are predominantly located in the region around the magic numbers. Nuclei with proton and neutron numbers situated between two magic numbers tend to exhibit larger deformations. Additionally, for nuclei located in the middle of the major shells, deformations in the light and medium-mass regions are generally more pronounced than those in heavy and superheavy nuclei.

For the neutron magic numbers $N$=8, 20, 50, 82, and 126, the corresponding isotonic chains are predominantly composed of spherical nuclei. However, the isotonic chain with neutron number $N$=28 exhibits significant shape shifts. Isotopic chains characterized by proton magic numbers ($Z$=8, 20, 50, 82) generally maintain spherical configurations, whereas the $Z$=28 isotopic chain exhibits moderate quadrupole deformation, which is not as pronounced as that in the isotonic chain with $N$=28.

For nuclei with neutron numbers $Z$>50, those located in the central region between major nuclear shells predominantly exhibit prolate deformation. A characteristic shape transition occurs as the neutron number approaches magic numbers, where nuclear deformation evolves from prolate to oblate configurations. Within the proton number range of 28<$Z$<50, oblate nuclear configurations are predominantly found near the central region of the major shells. Similarly, the range of 8<$Z$<20 also contains oblate nuclei. However, according to the IQMDD3 model, no significantly deformed oblate nuclei are predicted to exist within the 20<$Z$<28 range.\\

\subsection{\label{sec:leve33}Charge radii}
\begin{figure}[htbp]
	\centering
	\includegraphics[width=0.49\textwidth]{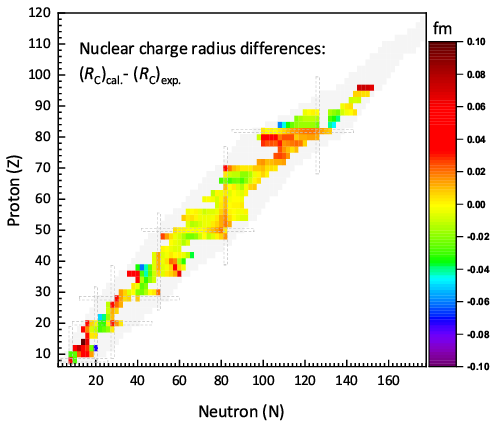}
	\caption{Discrepancies between IQMDD3+RHB calculations and experimental data for even-even nuclei with measured charge radii.}
	\label{fig:3}
\end{figure}

Besides nuclear deformations, the charge radius is also a crucial quantity for characterizing finite nuclei, and is expressed as: 
\begin{eqnarray}
	R_c = \sqrt{R^2_p+0.64} ~~~(\mathrm{fm}),
\end{eqnarray}
where the factor 0.64 corrects for the finite size of the proton charge distribution. The deviations between the theoretical values derived from the IQMDD3 model and the experimental data cited in Ref. \cite{Angeli2013,Tao2021} are depicted in Fig.~\ref{fig:3}. It shows that the IQMDD3 predictions are in reasonable agreement with the experimental charge radii, with calculated deviations typically within the $\pm0.04$ fm range. For the 368 even-even nuclei with known charge radii, the rms deviation between the theoretical and experimental values is $\sigma = 0.022$ fm, compared to $\sigma = 0.021$ fm for IQMDD-2*. 

The most significant discrepancies occur in the Mg($Z$=12), Si($Z$=14), and Cm($Z$=96) isotopic chains, where theoretical values exceed experimental values by over 0.05 fm. For $^{26}$Mg, $^{28}$Si, and $^{30}$Si, the theoretical overestimation of the charge radii relative to experimental values can be attributed to the significant oblate deformation ($|\beta_2|>0.2$) predicted in their ground states. Experiments reveal that the Cm ($Z$=96) isotopic chain exhibits anomalous charge radii, which are smaller than those of Pu ($Z$=94) \cite{Angeli2013}. However, IQMDD3 calculations indicate that charge radii typically increase with proton number in an isotonic chain. This results in a noticeable discrepancy between theoretical predictions and experimental observations for the charge radii of the Cm isotopic chain. Notably, this anomalous charge radius behavior exhibited by U-Pu-Cm isotopes has already been extensively discussed in Refs. \cite{Agbemava2014,Zhang2022,Guo2024}.

\subsection{\label{sec:leve34}Two-neutron and two-proton separation energies}
\begin{figure}[htbp]
	\centering
	\includegraphics[width=0.48\textwidth]{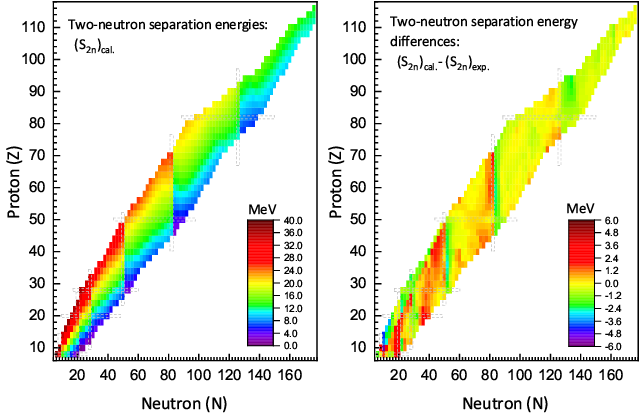}
	\caption{Left: The two-neutron separation energies of even-even nuclei calculated by IQMDD3. Right: The deviation between the theoretical and experimental values of the two-neutron separation energies.}
	\label{fig:4}
\end{figure}

Two-nucleon separation energies ($S_{2n}$ and $S_{2p}$) offer essential insights into the investigation of neutron and proton drip lines, nuclear shell evolution, and nucleon radioactivity phenomena \cite{Wang2025}. They can be expressed as:
\begin{eqnarray}
	S_{2n}(A,Z)&=&B(A,Z)-B(A-2,Z),\nonumber\\
	S_{2p}(A,Z)&=&B(A,Z)-B(A-2,Z-2),
\end{eqnarray}
where $B(A,Z)$ represents the binding energy of a nucleus with mass number $A$ and proton number $Z$, calculated from equation (16). In Fig.~\ref{fig:4}, the predicted two-neutron separation energies, $S_{2n}$, for even-even nuclei using the IQMDD3 parameters are depicted, along with the discrepancies between the theoretical predictions and experimental data from Ref. \cite{Wang2021}. The left panel of Fig.~\ref{fig:4} shows that, for a given isotopic chain, $S_{2n}$ decreases with increasing neutron number, while for a given isotonic chain, $S_{2n}$ increases with the increase in proton number. Meanwhile, as the proton number increases, the two-neutron separation energy $S_{2n}$ on the proton-rich side will gradually decrease. The theoretical values of $S_{2n}$ calculated by IQMDD3 are compared with the experimental values. As shown in the right panel of Fig.~\ref{fig:4}, most of the differences between theory and experiment remain within $\pm2$ MeV, and for the studied nuclei, the rms of the differences is determined to be 0.984 MeV. 

It is quite interesting to note that, as seen from the left panel of Fig.~\ref{fig:4}, there are pronounced sudden drops in $S_{2n}$ at the positions of $N$=50, 82 and 126. Similar sudden drops can also be observed at $N$=20 and 28. These abrupt decreases in $S_{2n}$ indicate the emergence of neutron shells, which are well-reproduced by IQMDD3.

\begin{figure}[htbp]
	\centering
	\includegraphics[width=0.48\textwidth]{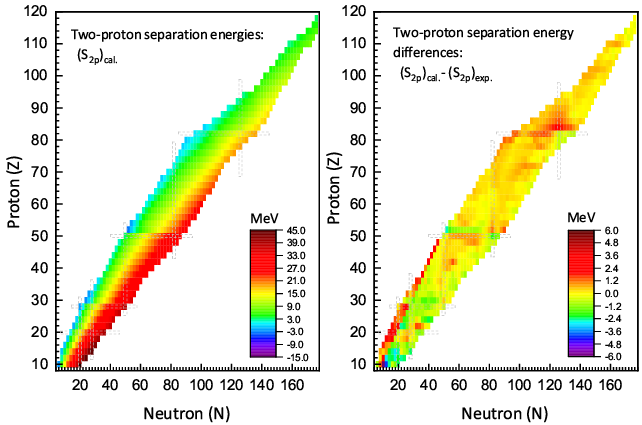}
	\caption{Left: Theoretical predictions of two-proton separation energies for even-even nuclei obtained from IQMDD3 calculations. Right: The discrepancy between theoretical calculations and experimental measurements of the two-proton separation energies.}
	\label{fig:5}
\end{figure}

Fig.~\ref{fig:5} presents the systematic trends of two-proton separation energy ($S_{2p}$) for even-even nuclei, accompanied by the differences between theoretical predictions and experimental measurements. In contrast to the two-neutron separation energy ($S_{2n}$), for a given isotopic chain, $S_{2p}$ exhibits an increasing trend with neutron number, while in isotonic chains, it demonstrates a decreasing trend with proton number. Furthermore, the data indicate a gradual reduction in $S_{2p}$ values within neutron-rich regions as the proton number increases. The right panel of Fig.~\ref{fig:5} demonstrates that the majority of theoretical-experimental deviations fall within a range of $\pm2$ MeV. For the nuclei under investigation, the rms deviation of $S_{2p}$ between theory and experiment is calculated to be 0.848 MeV, indicating acceptable agreement between theoretical predictions and experimental observations.

As observed in the left panel of Fig.~\ref{fig:5}, there are notable sudden drops in $S_{2p}$ at the proton numbers $Z$=20, 28 and 50. These abrupt decreases in $S_{2p}$ signify the appearance of proton shells. For the isotope chain with $Z$=82, the variation in $S_{2p}$ is relatively subtle. Nevertheless, the two-proton shell gaps, which will be discussed in the following subsection, provide a clearer manifestation of this change.

\subsection{\label{sec:leve35}Two-nucleon shell gaps}
The two-nucleon energy gaps, essential for examining the shell effects in finite nuclei, are defined as follows:
\begin{eqnarray}
	\delta_{2n}(A,Z)&=&S_{2n}(A,Z)-S_{2n}(A+2,Z),\nonumber\\
	\delta_{2p}(A,Z)&=&S_{2p}(A,Z)-S_{2p}(A+2,Z+2).
\end{eqnarray}

\begin{figure}[htbp]
	\centering
	\includegraphics[width=0.49\textwidth]{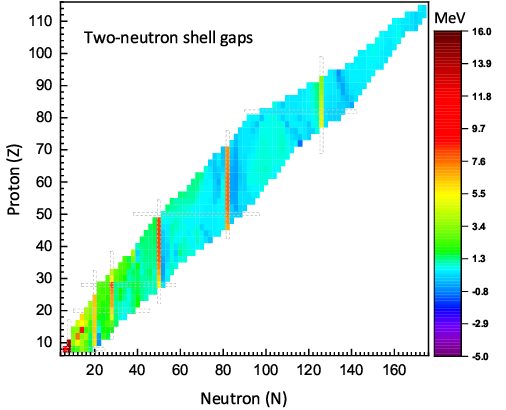}
	\caption{Two-neutron shell gaps $\delta_{2n}$ for even-even nuclei, as calculated with the IQMDD3 parameterization.}
	\label{fig:6}
\end{figure}

Fig.~\ref{fig:6} displays the two-neutron gaps ($\delta_{2n}$) for even-even nuclei in the region of $8\leq Z\leq114$. It is shown that $\delta_{2n}$ exhibits significant sharp changes at neutron numbers $N$=8, 20, 28, 50, 82, and 126, which can be regarded as one of the signatures of magic numbers. For the magic number $N$=28, $\delta_{2n}$ becomes relatively small in both proton-rich and neutron-rich regions, suggesting that the $N$=28 shell closure may be quenched in these areas.

\begin{figure}[htbp]
	\centering
	\includegraphics[width=0.49\textwidth]{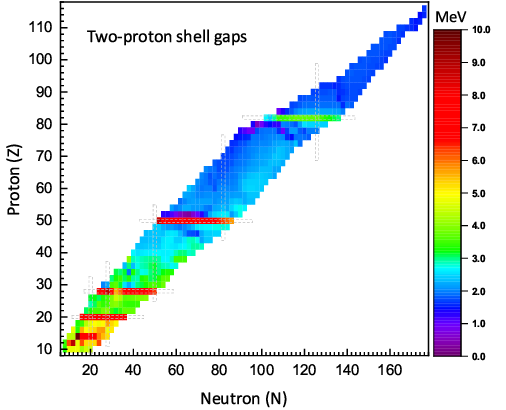}
	\caption{Two-proton shell gaps $\delta_{2p}$ in even-even nuclei from IQMDD3 calculations.}
	\label{fig:7}
\end{figure}

Fig.~\ref{fig:7} presents the two-proton gaps ($\delta_{2p}$) for even-even nuclei in the region of $10\leq Z\leq116$. It can be observed that $\delta_{2p}$ exhibits distinct peaks at proton numbers $Z$=20, 28, and 50, which is consistent with the conclusions regarding shell closures shown in Fig.~\ref{fig:5}. For $Z$=82, $\delta_{2p}$ also shows a significant sharp change, whereas the change in $S_{2p}$ is less evident in Fig.~\ref{fig:5}. This indicates that, compared to the two-proton separation energy, the two-proton gap provides a more intuitive indication of potential magic numbers. Additionally, it is noteworthy that a relatively large $\delta_{2p}$ is also observed at $Z$ = 14, suggesting that $Z$ = 14 may correspond to a subshell closure.

\subsection{\label{sec:leve36}$\alpha$-decay energies}
$\alpha$-decay usually occurs in heavy and superheavy radioactive nuclides. The $\alpha$-decay energy, denoted as $Q_{\alpha}$, is expressed as:
\begin{eqnarray}
	Q_{\alpha} = B(A-4,Z-2) + B(2,2) - B(A,Z).
\end{eqnarray}

\begin{figure}[htbp]
	\centering
	\includegraphics[width=0.49\textwidth]{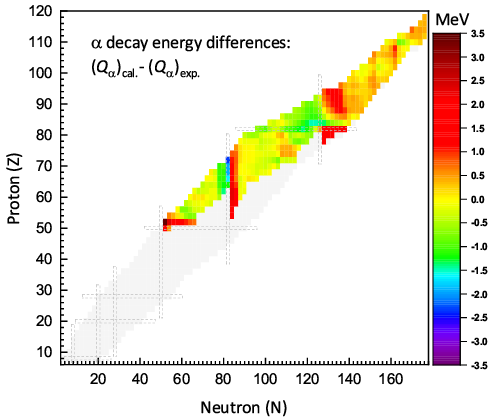}
	\caption{The discrepancies between the IQMDD3-calculated and experimental values of $\alpha$ decay energy $Q_{\alpha}$.}
	\label{fig:8}
\end{figure}

Fig.~\ref{fig:8} presents the discrepancies between the $Q_{\alpha}$ values calculated using the IQMDD3 model and the experimental data from Ref. \cite{Wang2021}. Among the 405 studied even-even nuclei with positive $Q_{\alpha}$ values, relatively larger deviations are observed in specific nuclei of the $Z$=52 and $Z$=94 isotopic chains, as well as in some nuclei along the $N$=84 isotonic chain. Overall, most of the discrepancies are within $\pm1.5$ MeV, and the rms of the discrepancies is 0.733 MeV. Recently, machine learning approaches have been successfully employed in the description of $\alpha$-decay energies, achieving a remarkably high level of precision \cite{Yuan2022,Yuan2024}. 

\section{Summary}
In this work, we have developed a relativistic Hartree-Bogoliubov (RHB) model incorporating quark-meson couplings and established a new parameter set named IQMDD3 by fitting to the binding energies and charge radii of twenty selected even-even nuclei. Using this parameter set, we systematically studied the ground-state properties of 868 even-even nuclei, with proton numbers from $Z$=8 to $Z$=118. For the twenty nuclei employed for fitting, the rms deviation of the binding energy per nucleon obtained with the IQMDD3 model is 0.023 MeV, that of the total binding energy is 2.33 MeV, and that of the charge radius is 0.01 fm. The predicted neutron skin thickness for $^{208}$Pb with this parameter set is 0.24 fm, with the symmetry energy and its slope at saturation density being 36.00 MeV and 100.1 MeV, respectively.

For the 868 even-even nuclei studied, the rms deviation of the binding energy per nucleon is 0.034 MeV, and the rms deviation of the total binding energy is 2.893 MeV. We investigated the systematic trends and general features of quadrupole deformations in these even-even nuclei. Results show that nuclei in the magic number regions and those in their vicinity are mostly spherical, except for the $N$=28 isotonic chain, which exhibits significant shape shifts. In contrast, nuclei far from the major shells exhibit larger deformations, with the largest deformations often occurring in the central regions between major shells. The analysis of charge radii shows that, for the 368 studied even-even nuclei with available experimental data, the rms deviation of the charge radii is 0.022 fm.

We systematically examined the two-nucleon separation energies, as well as the two-nucleon shell gaps. For the same isotopic chain, the two-neutron separation energy is larger in proton-rich regions than in neutron-rich regions, while the two-proton separation energy is the opposite, being larger in neutron-rich regions than in proton-rich regions. The conventional magic numbers were successfully reconstructed based on the systematic trends observed in two-nucleon separation energies and two-nucleon shell gaps. Additionally, the two-proton shell gaps indicate that $Z$ = 14 might be a subshell closure.

Finally, for nuclei with positive $\alpha$-decay energies, the extracted $Q_{\alpha}$ values were compared with the 405 available experimental data, yielding a rms deviation of $\sigma = 0.733$ MeV.

To further examine the validity and reliability of the IQMDD3 parameter set in describing finite nuclei, future work will employ this model to study the properties of drip-line nuclei far from the $\beta$-stability line.

\begin{acknowledgments} 

This work is supported by the National Natural Science Foundation of China (Grants No. 12535009, No. 11605093, and No. 12475135),
and by the Natural Science Youth Fund of Jiangsu Province (Grant No. BK20161036).
\end{acknowledgments}

\nocite{*}
\bibliography{bibliography}

\end{document}